\documentclass[%
 pra
 aip,
 jmp,%
 amsmath,amssymb,
twocolumn,superscriptaddress]{revtex4-2}

\usepackage{graphicx}
\usepackage{dcolumn}
\usepackage{bm}
\usepackage{hyperref}

\usepackage{booktabs}
\usepackage{physics}
\usepackage{xcolor}
\usepackage{color}
\usepackage{multirow}
\usepackage{float}

\newcommand{\vv}[1]{\boldsymbol{\mathbf{#1}}}
\usepackage{placeins}

\begin{document}
\preprint{APS/123-QED}

\title{Magnetic hyperfine structure constants of $^{137}$BaF in the $^2\Pi_{1/2}$ and $^2\Pi_{3/2}$ excited states }

\author{Yuly Chamorro}
\email{y.a.chamorro.mena@rug.nl}
\affiliation{Van Swinderen Institute for Particle Physics and Gravity, University of Groningen, 9747 AG, Groningen, The Netherlands}

\author{Felix Kogel}
\affiliation{5. Physikalisches  Institut  and  Center  for  Integrated  Quantum  Science  and  Technology, Universit\"at  Stuttgart,  Pfaffenwaldring  57,  70569  Stuttgart,  Germany}

\author{Tim Langen}
\affiliation{5. Physikalisches  Institut  and  Center  for  Integrated  Quantum  Science  and  Technology, Universit\"at  Stuttgart,  Pfaffenwaldring  57,  70569  Stuttgart,  Germany}
\affiliation{Vienna Center for Quantum Science and Technology, Atominstitut, TU Wien,  Stadionallee 2,  A-1020 Vienna,  Austria}

\author{Anastasia Borschevsky}
\affiliation{Van Swinderen Institute for Particle Physics and Gravity, University of Groningen, 9747 AG, Groningen, The Netherlands}

\date{\today}

\begin{abstract}
High-precision molecular experiments testing the Standard Model of particle physics require an accurate understanding of the molecular structure at the hyperfine level, both for the control of the molecules and for the interpretation of the results. In this work, we calculate the hyperfine structure constants for the excited states $^2\Pi_{1/2}$ and $^2\Pi_{3/2}$ of $^{137}$BaF due to the $^{137}$Ba nucleus. We use the 4-component relativistic Fock-space coupled-cluster method, extrapolating our results to the complete basis set limit. We investigate the effect of the basis sets and electron correlation, and estimate the uncertainty in our final results. Our results are used in the interpretation of the experimental spectroscopy of the hyperfine and rovibrational spectra of BaF, and the planning of laser-cooling schemes for future parity-violating anapole moment measurements \cite{KogChaLan25}.

\end{abstract}
\maketitle

\section{Introduction}
The hyperfine structure (HFS) of atoms and molecules arises as a result of the interaction of the electrons with the nuclear electromagnetic moments. Accurate determination of the hyperfine structure is important for understanding the underlying interactions at the atomic, nuclear, and particle physics level. On the one hand, the comparison of high-precision HFS measurements with theoretical predictions can be used to test nuclear models \cite{RobGin21,GinVol18} and quantum electrodynamics \cite{JenYer06,Kar05} and to search for slight variations in the fundamental constants \cite{FlamDzu09}. On the other hand, knowledge of the HFS of atoms and molecules is required to design and interpret various experiments on these systems.

In particular, knowledge of the HFS in molecules is needed in symmetry violation precision experiments searching for physics beyond the Standard Model of particle physics~\cite{Safronova2018,DeMille2023}. The success of these experiments relies on the precise control of the molecules in electric and magnetic fields.  Furthermore, the statistical sensitivity of many of these experiments benefits from the slowing of the molecular beams and the increase of their intensity by laser cooling \cite{BoeMarWil24,LimAlmHin18,KozHut17}. Designing laser cooling schemes requires knowledge of the complex molecular structure, including the HFS~\cite{HaoPasVis19,KogGarLan25serro}. For many systems, the HFS has not been experimentally measured, which motivates the use of electronic structure methods to predict it.

BaF is used in symmetry violation experiments that search for the electron electric dipole moment (eEDM) \cite{BoeMarWil24} and the nuclear-spin-dependent parity-violating anapole moment \cite{AltEmiDeM18,KogChaLan25,KogGarLan25} (NSD-PV). The experiments that search for the  eEDM use the even isotopologue of BaF, $^{138}\rm{Ba}{F}$, for which the HFS (due to the $^{19}$F atom) of the $^2\Sigma_{1/2}$ ground and $^2\Pi$ excited states has been measured and calculated~\cite{ErnKanTor86,DenHaaYin22}. NSD-PV searches use the odd isotopologue of BaF, $^{137}\rm{Ba}{F}$. While the HFS of $^{137}\rm{Ba}{F}$, due to $^{137}$Ba atom, in the $^2\Sigma_{1/2}$ ground state has been measured \cite{RylSchTor82} and calculated \cite{HaaEliBor20}, the HFS of the low-lying $^2 \Pi$ excited states has not been studied so far, neither experimentally nor theoretically.
The knowledge of the HFS of these states is necessary for the interpretation of the measured spectrum of $^{137}\rm{Ba}{F}$ and for the design of laser cooling schemes for future measurements of the parity violating anapole moment \cite{KogGarLan25, Langen2024}. 

In this work, we provide the theoretical predictions of the HFS constants for the $^2 \Pi$ excited states of $^{137}\rm{Ba}{F}$. The transition energies between the $^2\Sigma_{1/2}$ ground state and the $^2\Pi_{1/2}$ and $^2\Pi_{3/2}$ excited states are 11647 and 12273 cm$^{-1}$, respectively, which is convenient for laser cooling \cite{HaoPasVis19,KogRocLan21}.
Accurate calculations of the HFS require careful treatment of relativistic and electron-correlation effects. Relativistic effects are important in the calculation of molecular properties of systems containing heavy atoms \cite{pyy88} such as Ba and are especially important for the HFS constants in both heavy and light atoms, as errors of about 4\% have been found due to non-relativistic treatment of HFS constants even in light elements such as Ca \cite{PenSal84}. High-level treatment of electron correlation is not less important for achieving accurate and reliable predictions of these parameters \cite{HaaEliBor20,DenHaaYin22}.  For atoms, accurate calculations of hyperfine structure of the ground and excited states are routinely performed using various approaches that account for both relativistic and electron correlation effects \cite{RaeAckBac18, PorCheSaf22, GusRicRei20}. In contrast, for molecules, a limited number of such investigations have been performed so far, using multireference configuration interaction approach \cite{FleNay14} and single and multireference coupled cluster methods \cite{HaaEliBor20,DenHaaYin22,OleSkrSha20}. In particular, previous calculations of the HFS of $^{138}$BaF and $^{137}$BaF ($^2\Sigma_{1/2}$ ground state) using the relativistic Fock-space coupled-cluster method have shown a very good agreement with the experimental results \cite{HaaEliBor20,DenHaaYin22}. 
In this work, we follow a similar approach to the one used in Refs.~\cite{HaaEliBor20,DenHaaYin22}. We use the relativistic Fock space coupled cluster method and investigate the effect of the basis sets, electron correlation, and vibrational corrections on the calculated HFS constants. Using this computational study and the scheme derived in Refs. \cite{GusRicRei20, HaaEliBor20} we assess the uncertainty on our final results. 
\section{Theory}
The Hamiltonian describing the hyperfine interaction due to a nuclear spin $\vv{I}$ is given by, 
\begin{equation}
    H_{\mathrm{hfs}}=-\frac{ce\mu_0\gamma \vv{I}}{4\pi}
    \cdot
    \frac{ \vv{\alpha}\times \vv{r}}{r^3},
\end{equation}
where $c$ is the speed of light, $e$ the electric charge of the electron, $\mu_0$ de vacuum permeability, $\gamma$ the gyromagnetic ratio of the nucleus, $\vv{r}$ the position vector between the electron and the nucleus, and $\vv{\alpha}$ are the Dirac matrices \cite{Dya07book}. 
%
The hyperfine coupling tensor is defined as
\begin{equation}
 \vv{A}=\frac{1}{\Omega}\left\langle\frac{d  H_{\mathrm{hfs}}}{d \vv{I}}\right\rangle,
\end{equation}
with its parallel ($A_\parallel$) and perpendicular ($A_\perp$) components given by
\begin{equation}
 A_{z}=-\frac{1}{\Omega}\frac{ce\mu_0\gamma}{4\pi}
    \left\langle\frac{ (\vv{\alpha}\times \vv{r})_{z}}{r^3}\right\rangle=A_{\parallel},
\end{equation}
\begin{equation}
 A_{x/y}=-\frac{1}{\Omega}\frac{ce\mu_0\gamma}{4\pi}
    \left\langle\frac{ (\vv{\alpha}\times \vv{r})_{x/y}}{r^3}\right\rangle=A_{\perp},
\end{equation}
where $\Omega$ is the projection of the electronic angular momentum on the molecular axis.

We calculate the parallel and perpendicular hyperfine constants for the $^2\Pi_{1/2}$ and $^2\Pi_{3/2}$ excited states of $^{137}$BaF using the relativistic four-component Dirac-Coulomb Hamiltonian combined with the Fock-space coupled cluster method and the finite field approach, as described in detail in Ref. \cite{HaaEliBor20}. 
In this approach, the total Hamiltonian is expressed as a sum of the unperturbed Hamiltonian plus a perturbation regulated by the parameter $\lambda$, i.e. $H=H_0+\lambda \tilde{H}_{\mathrm{hfs}}$. We use the unperturbed Dirac-Coulomb Hamiltonian $H_0$,

\begin{equation}
    H_0=\sum_i[\beta_imc^2 + c \vv{\alpha}_i \cdot \vv{p}_i-V_{\text{nuc}}(\vv{r}_i)],
\end{equation}
where $\vv{\alpha}_i$ and $\beta_i$ are the $4\times4$ Dirac matrices, $\vv{p}_i$ is the momentum of the electron $i$, and $V_{\text{nuc}}$ is the Coulomb potential energy at the position of the electron with respect to the considered nucleus. To obtain $\vv{A}$, we take the perturbation $\tilde{H}_{\mathrm{hfs}}$ 
\begin{equation}\label{eq:H_tilde}
    \tilde{H}_{\mathrm{hfs}}=-\frac{ce\mu_0\gamma}{4\pi}\frac{\vv{\alpha}\times\vv{r}}{r^3},
\end{equation}
and use the finite field method to obtain
\begin{equation}\label{eq:A}
 A_{\parallel/\perp}=\frac{1}{\Omega}\frac{d E_{\rm{CC}}^{z/x}(\lambda)}{d\lambda}.
\end{equation}
$E_{\rm{CC}}^{z/x}$ are the energies calculated with the coupled-cluster method, for the perturbation strength $\lambda$,  with the molecule either oriented along $z$ which is parallel to the spin-quantization axis, or along $x$, which is perpendicular to the spin-quantization axis.

\section{Computational details}
We calculated the hyperfine structure constants for the A $^2\Pi_{1/2}$ and A $^2\Pi_{3/2}$ excited states of $^{137}$BaF due to the $^{137}$Ba nucleus using the $\mathrm{DIRAC19}$ code \cite{DIRAC19,Sau20}. 
We used the experimental bond length of the $^2\Pi$ excited state of BaF, i.e. 2.183 \AA~ \cite{Hub13}, and the nuclear magnetic dipole moment of $^{137}$Ba ($I=3/2$), i.e. $\mu(^{137}\text{Ba})=0.9375\mu_N$ \cite{Sto19,AntRodJas13}. 

The relativistic four-component Dirac-Coulomb Hamiltonian combined with the Fock-space coupled-cluster (FSCC) method \cite{VisEliKal01} and the finite field approach was used for the calculations. We used the perturbation presented in \autoref{eq:H_tilde} and obtained the HFS constants from the numerical derivative of the energy with respect to the perturbation, \autoref{eq:A}. We used the two-point formula to calculate the derivative, such that $A_{\parallel/\perp}$ is calculated as
\begin{equation}
    A_{\parallel/\perp}=\frac{1}{\Omega}\frac{E_{\rm{CC}}^{z/x}(\lambda) - E_{\rm{CC}}^{z/x}(-\lambda)}{2\lambda}.
\end{equation}
This approach is similar to the one employed in Ref. \cite{HaaEliBor20}, where $A_{\parallel/\perp}$ for the $^2\Sigma_{1/2}$ ground state of $^{137}$BaF were investigated and Ref. \cite{DenHaaYin22} where $A_{\parallel/\perp}$ for the $^2\Sigma_{1/2}$, $^2\Pi_{3/2}$ and $^2\Pi_{1/2}$ states in $^{138}$Ba$^{19}$F were calculated. 

Within the FSCC method, we used the (0h,1p) sector, i.e. we started the calculations from the singly ionized closed shell system and added a single electron within a model space constructed from the 6 lowest Kramers pairs, yielding the ground and the two lowest excited states of neutral BaF. We correlated all the electrons and included a virtual cut-off up to 2000 a.u., if not stated otherwise. We used the uncontracted relativistic Dyall basis sets \cite{Dya09,Dya12,Dya16} of double-, triple-, and quadruple-$\zeta$ cardinality, including correlating functions for the valence region (dyall.v$n\zeta$), for the valence and core region (dyall.cv$n\zeta$), or for all electrons (dyall.ae$n\zeta$) and also evaluated the effect of additional diffuse functions. We used the EXP-T code \cite{OleZaiEli20} to evaluate the effect of full triple excitations on the HFS constants. Furthermore, we calculated the vibrationally averaged HFS constants using the VIBCAL utility program in the DIRAC19 code.

\section{Results and discussion}
We investigated the effects of the quality of the basis sets (\autoref{sec:basis}) and electron correlation (\autoref{sec:correlation}) on the calculated $A_{\parallel/\perp}$ HFS constants for the $^2\Pi_{1/2}$ and $^2\Pi_{3/2}$ excited states of $^{137}$BaF. 
Based on the size of the effect of the various computational parameters on the calculated HFS constants, we determined the optimal scheme for the final recommended values; these were obtained by correlating all the electrons and using the Dyall v$n\zeta$ basis sets with $n=3,4$ \cite{Dya09,Dya12,Dya16} and extrapolating the values to the complete basis set (CBS) limit  (\autoref{sec:uncertainty}). We corrected these results by missing basis sets and vibrational effects (\autoref{sec:vib}) and estimated the uncertainty associated with our computational approach (\autoref{sec:uncertainty}). 


\subsection{Basis set quality}\label{sec:basis}
\autoref{tab:cardinality_basis} presents the obtained HFS constants for the dyall.v$n$z; $n=2,3,4$ basis sets, correlating all the electrons and using a virtual cut-off of 2000 a.u. The calculated HFS constants depend significantly on the cardinality of the basis sets, especially for the $^2\Pi_{3/2}$ state. 

\begin{figure}
    \centering
    \includegraphics[scale=0.7]{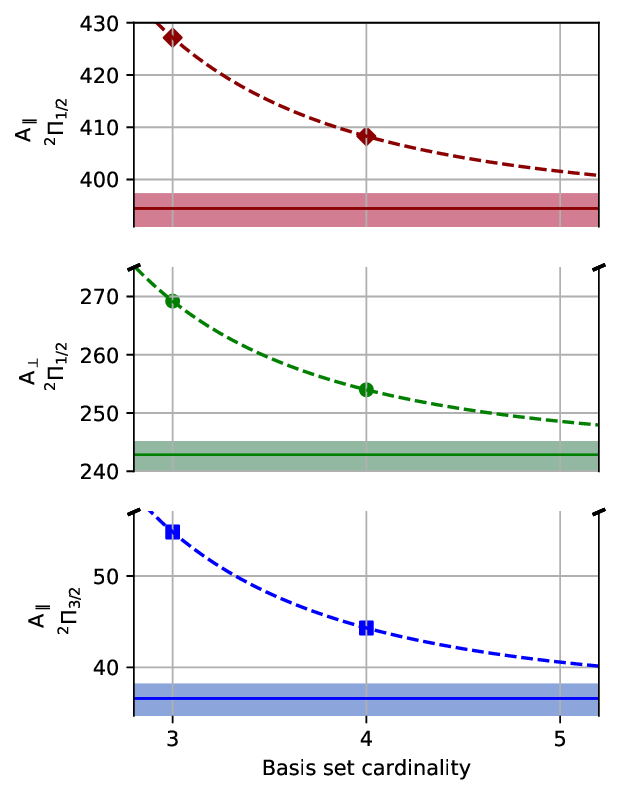}
    \caption{HFS constants (MHz) as a function of the cardinality of the dyall.v$n$z basis set, $n=3,4$. The dashed lines represent the curve fitted through the calculated HFS values and the CBS limit obtained using the Helgaker scheme; the latter is shown by the solid line. The shadowed interval is limited by the upper and lower CBS(M) and CBS(L) limits. }
    \label{fig:CBS}
\end{figure}

The HFS constants increase when going from $n=2$ to $n=3$, and decrease from $n=3$ to $n=4$. However, the energies in all the cases decrease with the cardinality of the basis sets. We neglected the less accurate $n=2$ results and used the $n=3,4$ results to calculate the CBS limit. We used the $n^{-3}$ ($n=3,4$) scheme from Helgaker et al. (CBS(H)) \cite{HelKloNog97} for extrapolating the energies and calculated the HFS constants via the finite field approach from these extrapolated energies. Furthermore, we tested the Martin $(n+\frac{1}{2})^{-4}$ scheme \cite{Mar96} and the scheme of Lesiuk and Jeziorski (based on the application of the Riemann zeta function) \cite{LesJez19}. Similarly to what we have found in our previous studies on symmetry-violating properties \cite{ChaBorPas22,ChaFlaPas24}, the limits obtained with the three schemes are evenly spread and give similar CBS limits, with a spread no larger than 6 MHz, within a 95\% confidence interval ($1.96\sigma$). \autoref{fig:CBS} shows the HFS constants as a function of the cardinality of the basis set together with the CBS limits. The calculation of the $A_{\parallel}$ constant for the $^2\Pi_{3/2}$ excited state (plot in blue) has the largest (relative) dependency on the cardinality of the basis set, with errors up to 50$\%$ for $n=3$ and 21$\%$ for $n=4$ when compared to the CBBS(H) limit. The errors on the $A_{\parallel}$ and $A_{\perp}$ constants for the $^2\Pi_{1/2}$ state are up to 11$\%$. 

\begin{table}[h]
\centering
\caption{Hyperfine constants (MHz) for increasing basis set quality and CBS limit using the Marten (M) \cite{Mar96}, Helgaker (H) \cite{HelKloNog97}, and Lisiuk-Jeziorski (L) \cite{LesJez19} schemes. The relative difference (\%) with respect to CBS(H) is presented in parentheses.}
\begin{tabular}{llll}
\hline
& \multicolumn{2}{c}{$^2\Pi_{1/2}$} &$^2\Pi_{3/2}$ \\
 & $A_{\parallel}$ & $A_{\perp}$ & $A_{\parallel}$ \\
\multicolumn{4}{l}{Cardinality} \\
\hline
v2z & 384.2 ($+$2.6) & 254.7 ($-$4.9) & 61.6 ($-$68)\\
v3z & 427.1 ($-$8.3) & 269.2 ($-$10.9) & 54.8 ($-$50)\\
v4z & 408.2 ($-$3.5) & 253.9 ($-$4.6) & 44.3 ($-$21)\\ 
\hline
CBS(M) & 397.3 ($-$0.7) &  245.1 ($-$1.0)& 38.2 ($-$4.4) \\
CBS(H) & 394.5        &  242.8 & 36.6 \\ 
CBS(L) & 391.0 ($+$0.9) &  240.0 ($+$1.2)& 34.7 ($+$5.3) \\
95\%c.i. & 6.2 &    5.0 &  3.5  \\
\hline
\end{tabular}\label{tab:cardinality_basis}
\end{table}

\autoref{tab:difuse_core_basis}
presents the effect of the missing core-correlating functions by comparing the dyall.v$3$z and the dyall.ae3z basis sets results. In these calculations, all electrons were correlated and a virtual cutoff of 2000 a.u. was used. The effect of additional diffuse functions was studied using the dyall.v$4z$ basis sets and correlating all the electrons for $A_{\parallel}$, but freezing the inner electrons in the calculation of $A_{\perp}$. The calculation of $A_{\perp}$ has a higher computational cost due to the restriction in symmetry, and here we limit our calculations to correlating 17 electrons.
The core correlating and diffuse functions have opposite effects on the HFS constants, with both contributions smaller than the effect of the basis set cardinality. 
Adding core-correlating functions leads to increase of the calculated the HFS constants by up to 2.9$\%$ (except for the $^2\Pi_{3/2}$ where we observe a negligible decrease of  0.5$\%$), while additional diffuse functions lead to a decrease of up to 5$\%$ for the highest lying $^2\Pi_{3/2}$ state. 
\begin{table}[h]
    \centering
    \caption{Effect of core-correlating and diffuse functions on the calculation of the HFS constants (in MHz). The relative percentages of difference with respect to the all-electron and augmented basis set results are presented in parentheses.}
        \begin{tabular}{llll}
        \hline
        & \multicolumn{2}{c}{$^2\Pi_{1/2}$} &$^2\Pi_{3/2}$ \\
        & $A_{\parallel}$ & $A_{\perp}$ & $A_{\parallel}$ \\
        \multicolumn{4}{l}{Core corr.} \\
        \hline
        dyall.v3z & 427.1 ($+$2.9) & 269.2 ($+$1.7) & 54.8 ($-$0.5)\\
        dyall.ae3z & 439.7 & 273.9 & 54.6 \\
        \multicolumn{4}{l}{Diffuse} \\
        \hline
        dyall.v4z & 408.2 ($-$1.8) & 229.7 ($-$3.3) & 44.3 ($-$5.0)\\
        dyall.v4z-s-aug & 401.0 & 222.5 & 42.2 \\
        \hline
    \end{tabular}\label{tab:difuse_core_basis}
\end{table}
We used the Helgaker CBS(H) limits and corrected them with the contributions due to the missing core-correlating and diffuse functions to obtain our final results.

\subsection{Electron correlation}\label{sec:correlation}

The calculated HFS constants discussed in the previous section were obtained within the FSCC method, including single and double excitations (FSCCSD), as implemented in the DIRAC19 code \cite{DIRAC19,VisEliKal01}. To evaluate the effect of the truncation of the coupled-cluster expansion, we used the FSCC method with up to triple excitations (FSCCSDT), implemented in the EXP-T code \cite{OleZaiEli20}. 
\autoref{tab:correlation} presents the results obtained by correlating 17 electrons and using the dyall.v3z basis set for the calculation of $A_{\parallel}$, and the dyall.v2z basis set for the calculation of $A_{\parallel}$ and $A_{\perp}$. It can be seen that the effect of the triple excitations on $A_{\parallel}$ is similar when using the 2z and the 3z basis sets; therefore, we used the 2z basis set for the more expensive calculation of the $A_{\perp}$ component. Inclusion of triple excitations leads to a decrease in the calculated HFS constants; similar behaviour was observed for the HFS constants of the $^2\Sigma_{1/2}$ ground state of $^{137}$BaF \cite{HaaEliBor20}. For the $^2\Pi_{1/2}$ state, $A_{\parallel}$ is significantly more sensitive to higher-order excitations than $A_{\perp}$. The effect of the triple excitations on $A_{\parallel}$ of the $^2\Pi_{3/2}$ state is much smaller. Due to the high computational costs that required use of larger basis sets we used these results as an estimate of the missing higher-order excitations in the uncertainty evaluation rather than as a correction in our final results.

Furthermore, we investigated the effect of the virtual space cut-off on the calculated HFS parameters. The results in \autoref{tab:correlation} show that the calculations of $A_{\parallel}$ and $A_{\perp}$ are converged at 2000 a.u. virtual cut-off. The effect of extending the cut-off to 3000 a.u. is negligible (no larger than 0.1$\%$). In these calculations, all electrons were correlated and the dyall.v3z basis sets were used. 

\begin{table}[]
\centering
\caption{Electron correlation effects on the calculated HFS constants (MHz). The differences with respect to to the FSCCSDT results or to the results obtained with the virtual space cut-off of 3000 a.u. are presented in parentheses.}
\begin{tabular}{llll}
\hline
& \multicolumn{2}{c}{$^2\Pi_{1/2}$} &$^2\Pi_{3/2}$ \\
 & $A_{\parallel}$ & $A_{\perp}$ & $A_{\parallel}$ \\
\multicolumn{4}{l}{Triple excitations} \\
\hline
FSCCSD-v3z & 367.9 ($-$1.5)  & -- & 51.2 ($-$0.4)\\
FSCCSDT-v3z & 362.5 & -- & 51.0 \\
FSCCSD-v2z & 347.3 ($-$1.6) & 229.8 ($-$0.6) & 56.6 ($+$0.3) \\
FSCCSDT-v2z & 342.0 & 228.5 & 56.8 \\ 
\multicolumn{4}{l}{Virtual space cutoff} \\
\hline
2000 a.u. & 427.1 ($+$0.02) &269.2 ($+$0.1) & 54.80 ($-$0.1) \\
3000 a.u. & 427.2&269.5 &54.75\\
\hline
\end{tabular}\label{tab:correlation}
\end{table}
~
\\

\subsection{Vibrational effects}\label{sec:vib}

\begin{figure}
    \centering
    \includegraphics[scale=0.6]{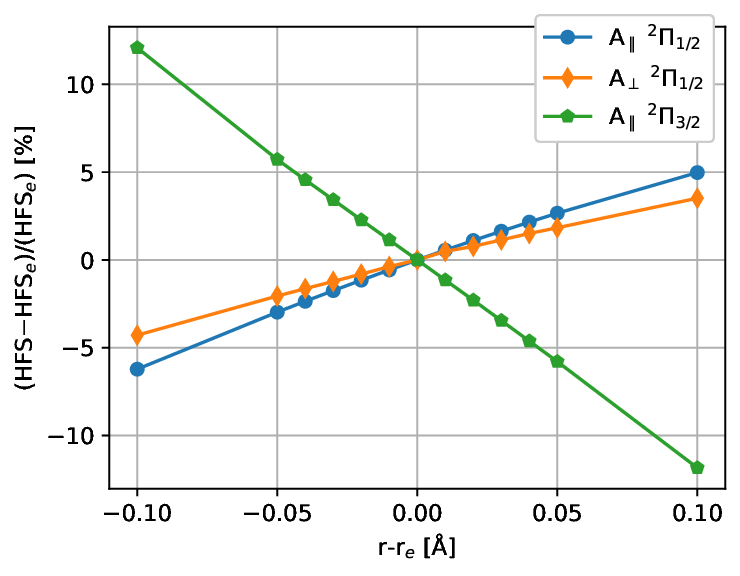}
    \caption{Variation (\%) of the HFS constants relative to the values at the equilibrium bond length ($r_e$) as a function of $\Delta r$.}
    \label{fig:distance}
\end{figure}

\begin{table}[h]
    \centering
    \caption{HFS constants (MHz) at different displacements ($\Delta r$) with respect to the equilibrium bond length ($r_e$). The vibrationally averaged HFS constants for the first vibrational state are also presented in the last line ($r_\nu=0$). }
        \begin{tabular}{rlll}
        \hline
        $\Delta r$ [\AA] & \multicolumn{2}{c}{$^2\Pi_{1/2}$} &$^2\Pi_{3/2}$ \\
        & $A_{\parallel}$ & $A_{\perp}$ & $A_{\parallel}$ \\
        \hline
-0.10 & 358.4 & 232.4 & 53.7 \\
-0.05 & 370.8 & 237.8 & 50.7 \\
-0.04 & 373.2 & 238.8 & 50.1 \\
-0.03 & 375.5 & 239.8 & 49.6 \\
-0.02 & 377.8 & 240.8 & 49.0 \\
-0.01 & 380.0 & 241.9 & 48.5 \\
0.00 & 382.2 & 242.8 & 48.0 \\
0.01 & 384.3 & 244.0 & 47.4 \\
0.02 & 386.4 & 244.7 & 46.8 \\
0.03 & 388.4 & 245.6 & 46.3 \\
0.04 & 390.4 & 246.4 & 45.7 \\
0.05 & 392.4 & 247.2 & 45.1 \\
0.10 & 401.2 & 251.3 & 42.2 \\
\hline
$r_{\nu=0}$ & 382.1 & 242.6  & 47.9\\

\hline
\end{tabular}\label{tab:distance}
\end{table}

We investigated the dependence of the HFS constants on the internuclear distance and subsequently calculated the vibrationally averaged HFS constants. For this analysis, we used the dyall.v3z basis set and correlated 17 electrons. We set a symmetric virtual space cut-off of 2 a.u.. We determined the equilibrium bond length at this level of theory using the TWOFIT utility program in the DIRAC19 code. We calculated the HFS constants at the equilibrium bond length ($r_e$), and at smaller and larger internuclear distances ($r$), i.e. at $r$, with $\Delta r=r-r_{e}$ = $\pm0.1, \pm0.05, \pm0.04, \pm0.03, \pm0.02, \pm0.01$ \AA. \autoref{tab:distance} presents the HFS constants as a function of the $\Delta r$ displacements, and \autoref{fig:distance} presents the variation of the calculated HFS constants with $\Delta r$ relative to the values obtained at the equilibrium bond length (HFS$_e$). The HFS parameter of the higher $^2\Pi_{3/2}$ state showed a stronger dependence and an opposite trend (decrease with increasing the bond length) compared to both HFS parameters of the $^2\Pi_{1/2}$ state. 
We used the VIBCAL utility in DIRAC19 to fit the bond-length dependence of the potential energies and HSF constants curves using a 6th-order polynomial. In the VIBCAL utility, the derivatives of the energy and the HFS constants were calculated and used to obtain the averaged properties for the first vibrational states.  
The vibrational corrections are found to be very small, as observed in \autoref{tab:distance} by comparing the HFS constants calculated at the equilibrium bond length ($\Delta r=0$) with the vibrationally averaged HFS constants of the first vibrational state ($r_{\nu=0}$). These corrections are 0.02\%, 0.12\%, and 0.28\% for the values of $A_{\parallel}$, $A_{\perp}$ for the $^2\Pi_{1/2}$ state, and $A_{\parallel}$ for the $^2\Pi_{3/2}$ state, respectively. 
A similar effect was observed in the molecular enhancement factors of the symmetry-violating electric dipole moment of the electron in BaF, where errors off 0.03\% and 0.15\% were found \cite{HaaDoeBoe21}.
We included this effect in our corrections and uncertainty estimation in the next section.

\subsection{Final results and uncertainty estimation}\label{sec:uncertainty}

Based on the computational study discussed in the previous sections, we chose the optimal approach for determining the final recommended values of the HFS constants and for assessing the uncertainty of these results.
The basis-sets cardinality has the largest effect on the calculated $A_{\parallel}$ and $A_{\perp}$ constants for both the $^2\Pi_{1/2}$ and $^2\Pi_{3/2}$ states. Therefore, we used the CBS(H) limit obtained with the dyall.v$n$z basis set as the baseline values for our final result. We corrected these values by the missing basis set  and vibrational effects, by using the values presented in \autoref{tab:results}. Specifically, the contributions from the core-correlating and additional diffuse functions missing in the baseline calculations were included by taking the difference between the dyall.ae3z and the dyall.v3z values, and the difference between the dyall.s-aug-v4z and the dyall.v4z results, respectively, (Table \ref{tab:difuse_core_basis}). The vibrational correction was obtained from the difference between the vibrationally averaged HFS constants and the HFS constants at the equilibrium bond length (Table \ref{tab:distance}).
Our final results, including the uncertainties (see discussion below) are compared to the experimental values \cite{KogChaLan25} in \autoref{tab:results}. 

\begin{figure}[]
    \centering
    \includegraphics[scale=0.65]{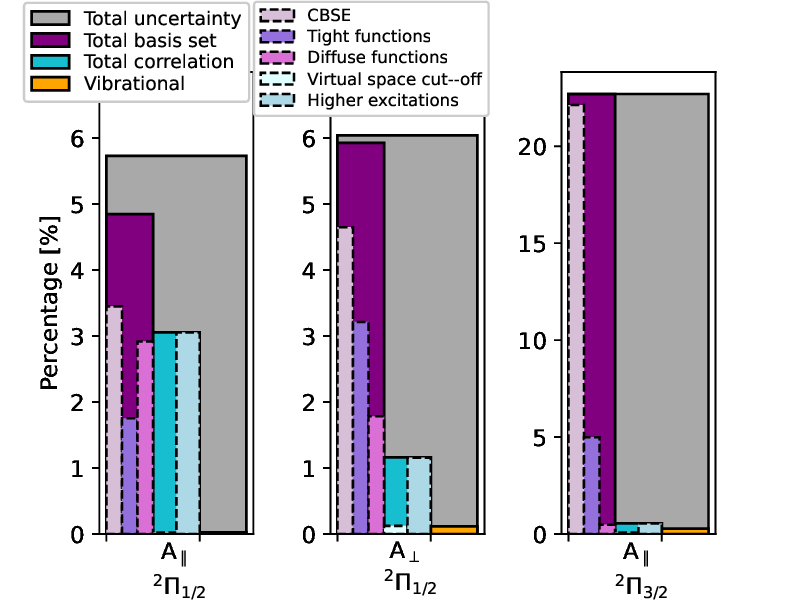}
    \caption{Contributions of the different sources to the total uncertainty on the final calculated HFS constants. The percentage is relative to the final value shown in \autoref{tab:uncertainty}.}
    \label{fig:uncertainty}
\end{figure}

Our evaluation of the uncertainty of this result is also based on the computational study we performed, where we take the size of the obtained corrections as a conservative estimate of the missing contributions; this approach was successfully employed for a variety of atomic and molecular properties \cite{HaaEliBor20,DenHaaYin22,GusRicRei20}. Since we are dealing
with higher-order effects, this procedure can be performed separately for each computational parameter.  To account for the uncertainty arising from the CBS extrapolation (rather than from using a truly complete basis set), we took the difference between the calculated dyall.v4z results and the extrapolated CBS(H) limit. 
Since the vibrational corrections were determined from the PECs obtained by correlating a reduced number of electrons, and by using the smaller dyall.v3z basis sets, we included the vibrational correction in our uncertainty estimation. 
Additionally, we took the difference between the results obtained with a virtual cut-off of 2000 a.u. and 3000 a.u. as the uncertainty from the incomplete virtual space. We estimated the effect of the truncation in the coupled cluster expansion by taking the difference between the FSCCSDT and the FSCCSD results shown in Table \ref{tab:correlation}. Relying on the separability of the higher-order effects one can assume that the uncertainty contributions stemming from different sources are independent to a large degree. Thus, the total uncertainty is obtained by adding the individual sources of uncertainty using the usual Euclidean norm. \autoref{tab:uncertainty} summarizes the contributions of each source of uncertainty, which are also  shown in \autoref{fig:uncertainty}. 

The quality of the basis sets has a larger effect than the missing electron correlation on the HFS parameters of both the $^2\Pi_{1/2}$ and the $^2\Pi_{3/2}$ states; particularly, the basis sets cardinality is the largest source of error and it is especially large for $A_{\parallel}$ in the $^2\Pi_{3/2}$ state. This is in line with our earlier study on the HFS parameters due to F in the excited states of BaF \cite{DenHaaYin22}.  
The present study does not account for the Breit interaction or higher-order QED effects. However, the Gaunt contribution was found to be negligible for the ground-state hyperfine parameters of $^{137}$BaF~\cite{HaaEliBor20} and are expected to be even smaller in the excited states. However, this study was performed on uncorrelated level and thus further investigations of the effect of these effects within an accurate computational approach will be the subject of a future work.  

\begin{table}[]
\centering
\caption{Baseline results, corrections, and final recommended HFS constants (MHz). The experimental HFS constants from Ref. \cite{KogChaLan25} are presented for comparison.}
\begin{tabular}{lrrrr}
\hline
 & \multicolumn{2}{c}{$^2\Pi_{1/2}$} &$^2\Pi_{3/2}$ & \\
            & $A_{\parallel}$ & $A_{\perp}$ & $A_{\parallel}$ & \\
\hline
 Baseline CBS(H) & 394.5 & 242.8 & 36.6 & \\
 \multicolumn{5}{l}{Corrections}\\
 Diffuse functions   & -7.0 & -7.7 & -1.7 & \\
 Tight functions     & 11.6 &  4.3 & -0.2 & \\
 Vibrational effects & -0.1 & -0.3 & -0.1 & \\
 \hline
 Final results              & $399.0\pm22.9$ & $239.1\pm14.4$& $34.6\pm7.9$ & \\
 \hline
 Experiment        & $413.2\pm~0.6$    & $254.3\pm~0.5$ & --   & \\
 \hline
    \end{tabular}
    \label{tab:results}
\end{table}

\begin{table*}[]
\centering
\caption{Summary of the contribution of the different sources of uncertainty in MHz and in percent relative  to the final results.}
\begin{tabular}{llrrrrrrl}
& & \multicolumn{4}{c}{$^2\Pi_{1/2}$} &\multicolumn{2}{c}{$^2\Pi_{3/2}$} & \\
\hline
& & \multicolumn{2}{c}{$A_{\parallel}$} & \multicolumn{2}{c}{$A_{\perp}$} & \multicolumn{2}{c}{$A_{\parallel}$} & \\
\textbf{}         & Source         & $\delta_i$   & \%  & $\delta_i$ & \%         & $\delta_i$  & \%            & Scheme                       \\
\hline
\multirow{3}{*}{basis set}          & CBSE & -13.8 & 3.4 & -11.1 & 4.6 & -7.7 & -22.1 & CBSE-v4z              \\
                  & valence region & -7.0   & 1.8  & -7.7 & 3.2  & -1.7     & 5.0                & s-aug-v4z – v4z              \\
                  & core region    & 11.6   & 2.9   & 4.3 & 1.8    & -0.2     & 0.5                & ae3z-v3z                     \\
\hline                  
\multirow{2}{*}{elec. corr.} & virt.act. space   & 0.1  & 0.03 & 0.3 & 0.1 & -0.03 & 0.1   & 3000-2000 a.u. v3z \\
                  & triples        & -12.2   & 3.1   &  -2.8 & 1.2   & 0.2      & 0.5                  & 2(CCSDT-CCSD)  \\
\hline                                                  
Vibrational             &                & -0.1   & 0.03    & -0.3  & 0.1   & -0.1    &  0.3                 & averaged-eq.             \\
\hline                                                  
Total             &                & 22.7   & 5.7    &  14.4 & 6.0   & 7.9      & 22.7                  & $\sqrt(\sum x_i^2)$ \\
Final value & \textbf{}      & 399.0  & \textbf{}  &   239.1          &  & 34.6     &                      & \textbf{}                   \\
\hline
\end{tabular}\label{tab:uncertainty}
\end{table*}

The current results for the HFS constants of the $^2 \Pi$ state are in very good agreement with the experimental values\cite{KogChaLan25}, in particular for $A_{\parallel}$. It should be noted that the measurements and the calculations were performed in parallel. Thus, the good performance of the relativistic Fock space coupled cluster approach in this case conforms the strong predictive power of this method for excited state  HFS constants in heavy systems.

\section{Conclusions}
We report the relativistic coupled cluster predictions of the $^{137}$Ba hyperfine structure constants for the $^2\Pi_{1/2}$ and $^2\Pi_{3/2}$ excited states of $^{137}$BaF. In our calculations we used the 4-component relativistic Fock space coupled cluster method and extrapolated our results to the complete basis set limit. We found that when all electrons are correlated, the largest source of error is the cardinality of the basis sets, i.e. the HFS constants are highly sensitive to the basis sets' cardinality, especially for the $A_{\parallel}$ constant of the $^2\Pi_{3/2}$ electronic state. We therefore calculated the CBS limit and used it as the basis for our final results. We corrected the CBS limit by the missing (smaller) basis set and vibrational effects. Furthermore, we found that the truncation in the coupled cluster expansion is another significant source of error. Finally, we assessed the uncertainty in our final results due to the incompleteness of our computational approach. 

Building on our theoretical results, the reported HFS constants have been investigated experimentally in a separate work using both fluorescence and absorption spectroscopy. The results show very good agreement with the theoretical values calculated in the present work, as detailed in Ref. \cite{KogChaLan25}. The combined knowledge thus obtained sets the stage for improved laser cooling of $^{137}$BaF~\cite{KogRocLan21,KogGarLan25}, which is expected to significantly enhance measurements of NSD-PV in this molecule~\cite{AltEmiDeM18}.

\section{Acknowledgments}
We thank the Center for Information Technology at the University of Groningen for their support and for providing access to the Peregrine and Hábrók high-performance computing clusters. We acknowledge the support from the Dutch Research Council, NWM (VI.Vidi.192.088). F.K. and T.L. acknowledge support from the European Research Council (ERC) under the European Union’s Horizon 2020 research and innovation program (Grant agreement No. 949431), the RiSC program of the Ministry of Science, Research and Arts Baden-W\"urttemberg and Carl Zeiss Foundation. This research was funded in whole or in part by the Austrian Science Fund (FWF) 10.55776/PAT8306623.
\bibliography{references} 

\begin{thebibliography}{10}

\bibitem{KogChaLan25}
F.~Kogel, Y.~Chamorro, M.~Bhattarai, M.~Rockenhäuser, T.~Garg, D.~DeMille, A.~Borschevsky, and T.~Langen, ``High-resolution spectroscopy of barium monofluoride: Odd isotopologues, hyperfine structure and isotope shifts,'' 2025.

\bibitem{RobGin21}
B.~M. Roberts and J.~S.~M. Ginges, ``Hyperfine anomaly in heavy atoms and its role in precision atomic searches for new physics,'' {\em Phys. Rev. A}, vol.~104, p.~022823, Aug 2021.

\bibitem{GinVol18}
J.~S.~M. Ginges and A.~V. Volotka, ``Testing atomic wave functions in the nuclear vicinity: The hyperfine structure with empirically deduced nuclear and quantum electrodynamic effects,'' {\em Phys. Rev. A}, vol.~98, p.~032504, Sep 2018.

\bibitem{JenYer06}
U.~D. Jentschura and V.~A. Yerokhin, ``Quantum electrodynamic corrections to the hyperfine structure of excited $s$ states,'' {\em Phys. Rev. A}, vol.~73, p.~062503, Jun 2006.

\bibitem{Kar05}
S.~G. Karshenboim, ``Precision physics of simple atoms: Qed tests, nuclear structure and fundamental constants,'' {\em Physics Reports}, vol.~422, no.~1, pp.~1--63, 2005.

\bibitem{FlamDzu09}
V.~V. Flambaum and V.~A. Dzuba, ``Search for variation of the fundamental constants in atomic, molecular, and nuclear spectra,'' {\em Canadian Journal of Physics}, vol.~87, no.~1, pp.~25--33, 2009.

\bibitem{Safronova2018}
M.~S. Safronova, D.~Budker, D.~DeMille, D.~F.~J. Kimball, A.~Derevianko, and C.~W. Clark, ``Search for new physics with atoms and molecules,'' {\em Rev. Mod. Phys.}, vol.~90, p.~025008, Jun 2018.

\bibitem{DeMille2023}
D.~DeMille, N.~R. Hutzler, A.~M. Rey, and T.~Zelevinsky, ``Quantum sensing and metrology for fundamental physics with molecules,'' {\em Nature Physics}, vol.~20, no.~5, pp.~741--749, 2024.

\bibitem{BoeMarWil24}
A.~Boeschoten, V.~R. Marshall, T.~B. Meijknecht, A.~Touwen, H.~L. Bethlem, A.~Borschevsky, S.~Hoekstra, J.~W.~F. van Hofslot, K.~Jungmann, M.~C. Mooij, R.~G.~E. Timmermans, W.~Ubachs, and L.~Willmann, ``Spin-precession method for sensitive electric dipole moment searches,'' {\em Phys. Rev. A}, vol.~110, p.~L010801, Jul 2024.

\bibitem{LimAlmHin18}
J.~Lim, J.~Almond, M.~Trigatzis, J.~Devlin, N.~Fitch, B.~Sauer, M.~Tarbutt, and E.~Hinds, ``Laser cooled ybf molecules for measuring the electron’s electric dipole moment,'' {\em Physical review letters}, vol.~120, no.~12, p.~123201, 2018.

\bibitem{KozHut17}
I.~Kozyryev and N.~R. Hutzler, ``Precision measurement of time-reversal symmetry violation with laser-cooled polyatomic molecules,'' {\em Physical review letters}, vol.~119, no.~13, p.~133002, 2017.

\bibitem{HaoPasVis19}
Y.~Hao, L.~F. Pašteka, L.~Visscher, P.~Aggarwal, H.~L. Bethlem, A.~Boeschoten, A.~Borschevsky, M.~Denis, K.~Esajas, S.~Hoekstra, K.~Jungmann, V.~R. Marshall, T.~B. Meijknecht, M.~C. Mooij, R.~G.~E. Timmermans, A.~Touwen, W.~Ubachs, L.~Willmann, Y.~Yin, A.~Zapara, and N.~eEDM Collaboration), ``High accuracy theoretical investigations of caf, srf, and baf and implications for laser-cooling,'' {\em The Journal of Chemical Physics}, vol.~151, p.~034302, 07 2019.

\bibitem{KogGarLan25serro}
F.~Kogel, T.~Garg, M.~Rockenhäuser, S.~A. Morales-Ramírez, and T.~Langen, ``Molecular laser cooling using serrodynes: implementation, characterization and prospects,'' {\em New Journal of Physics}, vol.~27, p.~055001, may 2025.

\bibitem{AltEmiDeM18}
E.~Altunta\ifmmode~\mbox{\c{s}}\else \c{s}\fi{}, J.~Ammon, S.~B. Cahn, and D.~DeMille, ``Demonstration of a sensitive method to measure nuclear-spin-dependent parity violation,'' {\em Phys. Rev. Lett.}, vol.~120, p.~142501, Apr 2018.

\bibitem{KogGarLan25}
F.~Kogel, T.~Garg, M.~Rockenh\"auser, and T.~Langen, ``Laser-cooled $^{137}\mathrm{BaF}$ molecules for measuring nuclear-spin-dependent parity violation,'' {\em Phys. Rev. Res.}, vol.~7, p.~L022041, May 2025.

\bibitem{ErnKanTor86}
W.~E. Ernst, J.~Kändler, and T.~Törring, ``Hyperfine structure and electric dipole moment of baf x$^2\sigma$+,'' {\em The Journal of Chemical Physics}, vol.~84, pp.~4769--4773, 05 1986.

\bibitem{DenHaaYin22}
M.~Denis, P.~A. Haase, M.~C. Mooij, Y.~Chamorro, P.~Aggarwal, H.~L. Bethlem, A.~Boeschoten, A.~Borschevsky, K.~Esajas, Y.~Hao, {\em et~al.}, ``Benchmarking of the fock-space coupled-cluster method and uncertainty estimation: Magnetic hyperfine interaction in the excited state of baf,'' {\em Physical Review A}, vol.~105, no.~5, p.~052811, 2022.

\bibitem{RylSchTor82}
C.~Ryzlewicz, H.-U. Schütze-Pahlmann, J.~Hoeft, and T.~Törring, ``Rotational spectrum and hyperfine structure of the 2$\sigma$ radicals baf and bacl,'' {\em Chemical Physics}, vol.~71, no.~3, pp.~389--399, 1982.

\bibitem{HaaEliBor20}
P.~A. Haase, E.~Eliav, M.~Ilias, and A.~Borschevsky, ``Hyperfine structure constants on the relativistic coupled cluster level with associated uncertainties,'' {\em The Journal of Physical Chemistry A}, vol.~124, no.~16, pp.~3157--3169, 2020.

\bibitem{Langen2024}
T.~Langen, G.~Valtolina, D.~Wang, and J.~Ye, ``{Quantum state manipulation and cooling of ultracold molecules},'' {\em Nature Physics}, vol.~20, no.~5, pp.~702--712, 2024.

\bibitem{KogRocLan21}
F.~Kogel, M.~Rockenhäuser, R.~Albrecht, and T.~Langen, ``A laser cooling scheme for precision measurements using fermionic barium monofluoride (137ba19f) molecules,'' {\em New Journal of Physics}, vol.~23, p.~095003, sep 2021.

\bibitem{pyy88}
P.~Pyykko, ``Relativistic effects in structural chemistry,'' {\em Chemical Reviews}, vol.~88, no.~3, pp.~563--594, 1988.

\bibitem{PenSal84}
A.-M. M\aa{}rtensson-Pendrill and S.~Salomonson, ``Hyperfine structure of the $4s$, $4p$, and $3d$ states in ${\mathrm{ca}}^{+}$ evaluated by many-body perturbation theory,'' {\em Phys. Rev. A}, vol.~30, pp.~712--721, Aug 1984.

\bibitem{RaeAckBac18}
S.~Raeder, D.~Ackermann, H.~Backe, R.~Beerwerth, J.~C. Berengut, M.~Block, A.~Borschevsky, B.~Cheal, P.~Chhetri, C.~E. D\"ullmann, V.~A. Dzuba, E.~Eliav, J.~Even, R.~Ferrer, V.~V. Flambaum, S.~Fritzsche, F.~Giacoppo, S.~G\"otz, F.~P. He\ss{}berger, M.~Huyse, U.~Kaldor, O.~Kaleja, J.~Khuyagbaatar, P.~Kunz, M.~Laatiaoui, F.~Lautenschl\"ager, W.~Lauth, A.~K. Mistry, E.~Minaya~Ramirez, W.~Nazarewicz, S.~G. Porsev, M.~S. Safronova, U.~I. Safronova, B.~Schuetrumpf, P.~Van~Duppen, T.~Walther, C.~Wraith, and A.~Yakushev, ``Probing sizes and shapes of nobelium isotopes by laser spectroscopy,'' {\em Phys. Rev. Lett.}, vol.~120, p.~232503, Jun 2018.

\bibitem{PorCheSaf22}
S.~G. Porsev, C.~Cheung, and M.~S. Safronova, ``Calculation of energies and hyperfine-structure constants of $^{233}\mathrm{U}^{+}$ and $^{233}\mathrm{U}$,'' {\em Phys. Rev. A}, vol.~106, p.~042810, Oct 2022.

\bibitem{GusRicRei20}
F.~P. Gustafsson, C.~M. Ricketts, M.~L. Reitsma, R.~F. Garcia~Ruiz, S.~W. Bai, J.~C. Berengut, J.~Billowes, C.~L. Binnersley, A.~Borschevsky, T.~E. Cocolios, B.~S. Cooper, R.~P. de~Groote, K.~T. Flanagan, A.~Koszor\'us, G.~Neyens, H.~A. Perrett, A.~R. Vernon, Q.~Wang, S.~G. Wilkins, and X.~F. Yang, ``Tin resonance-ionization schemes for atomic- and nuclear-structure studies,'' {\em Phys. Rev. A}, vol.~102, p.~052812, Nov 2020.

\bibitem{FleNay14}
T.~Fleig and M.~K. Nayak, ``Electron electric dipole moment and hyperfine interaction constants for tho,'' {\em Journal of Molecular Spectroscopy}, vol.~300, pp.~16--21, 2014.
\newblock Spectroscopic Tests of Fundamental Physics.

\bibitem{OleSkrSha20}
A.~V. Oleynichenko, L.~V. Skripnikov, A.~Zaitsevskii, E.~Eliav, and V.~M. Shabaev, ``Diagonal and off-diagonal hyperfine structure matrix elements in kcs within the relativistic fock space coupled cluster theory,'' {\em Chemical Physics Letters}, vol.~756, p.~137825, 2020.

\bibitem{Dya07book}
K.~G. Dyall and K.~F{\ae}gri~Jr, {\em Introduction to relativistic quantum chemistry}.
\newblock Oxford University Press, 2007.

\bibitem{DIRAC19}
{DIRAC}, a relativistic ab initio electronic structure program, Release {DIRAC19} (2019), written by A.~S.~P.~Gomes, T.~Saue, L.~Visscher, H.~J.~{\relax Aa}.~Jensen, and R.~Bast, with contributions from I.~A.~Aucar, V.~Bakken, K.~G.~Dyall, S.~Dubillard, U.~Ekstr{\"o}m, E.~Eliav, T.~Enevoldsen, E.~Fa{\ss}hauer, T.~Fleig, O.~Fossgaard, L.~Halbert, E.~D.~Hedeg{\aa}rd, B.~Heimlich--Paris, T.~Helgaker, J.~Henriksson, M.~Ilia{\v{s}}, Ch.~R.~Jacob, S.~Knecht, S.~Komorovsk{\'y}, O.~Kullie, J.~K.~L{\ae}rdahl, C.~V.~Larsen, Y.~S.~Lee, H.~S.~Nataraj, M.~K.~Nayak, P.~Norman, G.~Olejniczak, J.~Olsen, J.~M.~H.~Olsen, Y.~C.~Park, J.~K.~Pedersen, M.~Pernpointner, R.~di~Remigio, K.~Ruud, P.~Sa{\l}ek, B.~Schimmelpfennig, B.~Senjean, A.~Shee, J.~Sikkema, A.~J.~Thorvaldsen, J.~Thyssen, J.~van~Stralen, M.~L.~Vidal, S.~Villaume, O.~Visser, T.~Winther, and S.~Yamamoto (available at \url{http://dx.doi.org/10.5281/zenodo.3572669}, see also \url{http://www.diracprogram.org}).

\bibitem{Sau20}
T.~Saue, R.~Bast, A.~S.~P. Gomes, H.~J.~A. Jensen, L.~Visscher, I.~A. Aucar, R.~Di~Remigio, K.~G. Dyall, E.~Eliav, E.~Fasshauer, {\em et~al.}, ``The dirac code for relativistic molecular calculations,'' {\em The Journal of chemical physics}, vol.~152, no.~20, p.~204104, 2020.

\bibitem{Hub13}
K.~Huber, {\em Molecular spectra and molecular structure: IV. Constants of diatomic molecules}.
\newblock Springer Science \& Business Media, 2013.

\bibitem{Sto19}
N.~Stone, ``Table of recommended nuclear magnetic dipole moments,'' tech. rep., International Atomic Energy Agency, 2019.

\bibitem{AntRodJas13}
A.~Antušek, P.~Rodziewicz, D.~Ke¸dziera, A.~Kaczmarek-Ke¸dziera, and M.~Jaszuński, ``Ab initio study of nmr shielding of alkali earth metal ions in water complexes and magnetic moments of alkali earth metal nuclei,'' {\em Chemical Physics Letters}, vol.~588, pp.~57--62, 2013.

\bibitem{VisEliKal01}
L.~Visscher, E.~Eliav, and U.~Kaldor, ``Formulation and implementation of the relativistic fock-space coupled cluster method for molecules,'' {\em The Journal of Chemical Physics}, vol.~115, no.~21, pp.~9720--9726, 2001.

\bibitem{Dya09}
K.~G. Dyall, ``Relativistic double-zeta, triple-zeta, and quadruple-zeta basis sets for the 4s, 5s, 6s, and 7s elements,'' {\em The Journal of Physical Chemistry A}, vol.~113, no.~45, pp.~12638--12644, 2009.

\bibitem{Dya12}
K.~G. Dyall, ``Core correlating basis functions for elements 31--118,'' {\em Theoretical Chemistry Accounts}, vol.~131, pp.~1--11, 2012.

\bibitem{Dya16}
K.~G. Dyall, ``Relativistic double-zeta, triple-zeta, and quadruple-zeta basis sets for the light elements {H}--{Ar},'' {\em Theoretical Chemistry Accounts}, vol.~135, no.~5, p.~128, 2016.

\bibitem{OleZaiEli20}
A.~V. Oleynichenko, A.~Zaitsevskii, and E.~Eliav, ``Towards high performance relativistic electronic structure modelling: The exp-t program package,'' in {\em Supercomputing} (V.~Voevodin and S.~Sobolev, eds.), (Cham), pp.~375--386, Springer International Publishing, 2020.

\bibitem{HelKloNog97}
T.~Helgaker, W.~Klopper, H.~Koch, and J.~Noga, ``Basis-set convergence of correlated calculations on water,'' {\em The Journal of Chemical Physics}, vol.~106, no.~23, pp.~9639--9646, 1997.

\bibitem{Mar96}
J.~M. Martin, ``Ab initio total atomization energies of small molecules—towards the basis set limit,'' {\em Chemical Physics letters}, vol.~259, no.~5-6, pp.~669--678, 1996.

\bibitem{LesJez19}
M.~Lesiuk and B.~Jeziorski, ``Complete basis set extrapolation of electronic correlation energies using the {R}iemann zeta function,'' {\em Journal of Chemical Theory and Computation}, vol.~15, no.~10, pp.~5398--5403, 2019.

\bibitem{ChaBorPas22}
Y.~Chamorro, A.~Borschevsky, E.~Eliav, N.~R. Hutzler, S.~Hoekstra, and L.~F. Pa{\v{s}}teka, ``Molecular enhancement factors for the p, t-violating electric dipole moment of the electron in ${\mathrm{bach}_{3}}$ and ${\mathrm{ybch}_{3}}$ symmetric top molecules,'' {\em Physical Review A}, vol.~106, no.~5, p.~052811, 2022.

\bibitem{ChaFlaPas24}
Y.~Chamorro, V.~V. Flambaum, R.~F. Garcia~Ruiz, A.~Borschevsky, and L.~F. Pa{\v{s}}teka, ``Enhanced parity and time-reversal-symmetry violation in diatomic molecules: Lao, las, and luo,'' {\em Physical Review A}, vol.~110, no.~4, p.~042806, 2024.

\bibitem{HaaDoeBoe21}
P.~A. Haase, D.~J. Doeglas, A.~Boeschoten, E.~Eliav, M.~Ilia{\v{s}}, P.~Aggarwal, H.~L. Bethlem, A.~Borschevsky, K.~Esajas, Y.~Hao, {\em et~al.}, ``Systematic study and uncertainty evaluation of p, t-odd molecular enhancement factors in baf,'' {\em The Journal of chemical physics}, vol.~155, no.~3, 2021.

\end{thebibliography}
\bibliographystyle{ieeetr}

\end{document}